\documentclass[prd,twocolumn,nofootinbib,reprint]{revtex4-1}
\usepackage{amsmath,amssymb,amsthm,bm,braket,ascmac,bbm}
\usepackage{graphicx}
\usepackage{color} 

\theoremstyle{definition}

\begin{document}
\title{Inelastic Dark Matter Electron Scattering and the XENON1T Excess}
\author{Keisuke Harigaya$^1$, Yuichiro Nakai$^2$ and Motoo Suzuki$^2$}
\affiliation{\vspace{2mm} $^1$School of Natural Sciences, Institute for Advanced Study, Princeton, NJ 08540, USA \\
$^2$Tsung-Dao Lee Institute and School of Physics and Astronomy, \\Shanghai
Jiao Tong University, 800 Dongchuan Road, Shanghai, 200240 China \\}


\begin{abstract}
Detection of electron recoils by dark matter (DM)  may reveal the structure of the dark sector. We consider a scenario where a heavier DM particle inelastically scatters off an electron and is converted into a lighter DM particle. A small mass difference between the two DM particles is transferred into electron recoil energy. We investigate the DM-electron interaction mediated by a massive dark photon and evaluate the inelastic DM scattering rate, taking account of the atomic structure. It is found that the scattering rate is significantly enhanced because of the small mass splitting, which allows for a small momentum transfer matched with the size of the electron wave function.  We show that there exists a viable parameter space which explains the excess of electron recoil events around 2-3 keV recently reported by the XENON1T experiment.
\end{abstract}
\maketitle

\section{Introduction}\label{sec:intro}

To understand the nature of dark matter (DM) is a central issue in modern particle physics and cosmology.
Numerous candidates of DM have been proposed and at the same time
numerous experiments have been conducted to search for DM.
The dawn of a new era in DM physics is breaking. 
Recently, the XENON collaboration has reported excess of electron recoil events around 2-3 keV
in the recoil energy~\cite{Aprile:2020tmw}.
The observed excess was interpreted in terms of axions \cite{Peccei:1977hh,Weinberg:1977ma,Wilczek:1977pj}
produced in the Sun.
However, this interpretation is in strong tension with the stellar cooling constraints \cite{Viaux:2013lha,Bertolami:2014wua,Battich:2016htm,Giannotti:2017hny}.
Another interpretation based on a hypothetical neutrino magnetic moment (see e.g.~refs.~\cite{Fukugita:1987ti,Bell:2005kz,Bell:2006wi}) is also excluded by the same reason.
Then, barring the possibility that the signals come from a small amount of tritium in the detector,
it is natural to consider the excess as a hint of DM. 

The observed electron recoil excess cannot be explained by cold DM which elastically scatters off target electrons
because such DM particles are too slow and give too large signals in the first bin of the recoil energy 1-2 keV
when the second bin of 2-3 keV is fitted~\cite{Kannike:2020agf}.
One possible explanation of the excess is absorption of bosonic DM by electrons~\cite{Pospelov:2008jk}.
A concrete setup to realize this idea is discussed in ref.~\cite{Takahashi:2020bpq}.
Another explanation is to have a fast component of DM with velocity $v \sim 0.1$~\cite{Kannike:2020agf},
whose possible origins are also speculated. 

In this paper, we propose a new interpretation of the observed excess
with cold DM inelastically scattering off electrons. 
Inelastic DM scattering has been mostly discussed in the context of inelastic DM~\cite{Hall:1997ah,TuckerSmith:2001hy},
where a DM particle scatters off nuclei and is converted into an excited state,
motivated by the DAMA annual modulation anomaly~\cite{Bernabei:2000qi}.
Unlike inelastic DM, we consider a cold DM particle $\chi_2$ which inelastically scatters off an electron
and is converted into a lighter DM particle $\chi_1$. A DM nucleon down-scattering has been discussed in ref.~\cite{Dienes:2017ylr}.
The mass difference between $\chi_1$ and $\chi_2$ is converted into the electron recoil energy.

To be concrete, we investigate the DM-electron interaction mediated by a massive dark photon $A'$.
A DM particle $\chi_2$ can decay into  Standard Model (SM) particles and $\chi_1$,
but the lifetime is sufficiently long.
The inelastic scattering in the early universe freezes out much before the temperature drops below the mass difference.
The final abundance of the DM particle $\chi_2$ is somewhat suppressed compared to that of $\chi_1$
due to the annihilation of $\chi_2$ into $\chi_1$ after the temperature drops below the mass difference.
However, both of $\chi_1$ and $\chi_2$ are still important components of DM in the present universe.
We calculate the rate of the inelastic DM scattering off electrons, taking account of the xenon atomic structure. We find that the scattering rate is significantly enhanced for a recoil energy at the mass difference, since the momentum transfer is allowed to be small and can match the size of the wave function of the electrons in the atom. We find a viable parameter space of our dark sector model where the observed excess is explained.

The rest of the paper is organized as follows.
In section~\ref{sec:model}, we present our dark sector model.
In section~\ref{sec:excess}, the inelastic scattering cross section is computed.
Section~\ref{sec:FO} discusses the DM production through the thermal freeze out process.
In section~\ref{sec:constraints}, we investigate (in)direct constraints on the model.
The lifetime of the heavier DM component is estimated and the constraint from various dark photon searches is shown.
Then, in section~\ref{xenon1t}, we discuss the model parameter space which can explain the XENON1T data.
Section~\ref{sec:discussion} is devoted to conclusions and comments on future directions.

\section{The model}\label{sec:model}

We introduce a new sector with two DM scalars $\chi_1$, $\chi_2$
(whose masses are $m_{1} < m_{2}$) feebly interacting with the SM particles
through a massive dark photon $A'$.
Our focus is on the case where the DM masses $m_{1,2}$ are much above the MeV scale
so that the DM abundance may be explained by the thermal freeze-out process as discussed in section~\ref{sec:FO}. 
We assume a mass difference between $\chi_1$ and $\chi_2$,
defined as $\delta \equiv m_2 - m_1 \, (= 2-3 \, \rm keV)$, is much smaller than $m_{1,2}$.
Such a small mass difference can be understood by a gauge symmetry and its spontaneous breaking. We embed $\chi_{1,2}$ into a complex scalar field $\phi$ charged under a dark gauge symmetry $U(1)_D$.
To ensure the stability of the DM scalars, we impose a discrete symmetry $\phi \rightarrow -\phi$.
The $U(1)_D$ symmetry is spontaneously broken and the dark photon $A'$ becomes massive. 
The symmetry breaking generates a mass difference between $\chi_1$ and $\chi_2$ in $\phi$ through a potential \cite{Hall:1997ah,TuckerSmith:2001hy,Okada:2019sbb},
\begin{align}
    V(\phi) = m^2|\phi|^2 + \Delta^2 \left(\phi^2 + \phi^{*2} \right), 
\end{align}
where $\Delta$ originates from the gauge symmetry breaking.
Then, the masses of $\chi_{1,2}$ are calculated as
\begin{equation}
\begin{split}
    &m_1 = \sqrt{m^2- 2 \Delta^2} \simeq  m - \frac{\Delta^2}{m}, \\
    &m_2 = \sqrt{m^2+ 2 \Delta^2} \simeq  m + \frac{\Delta^2}{m},  \\
    &\delta = m_{2}- m_1 \simeq 2 \frac{\Delta^2}{m}.
\end{split}
\end{equation}
By taking $\Delta \ll m$, the mass difference $\delta$ is suppressed compared to $m_{1,2}$.

\begin{figure}[!t]
\begin{center}
  \includegraphics[width=6cm]{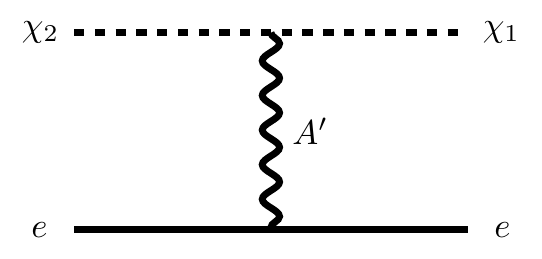}
 \end{center} 
  \vspace{-0.3cm}
  \caption{Inelastic scattering of the heavier DM particle $\chi_2$ off the electron $e$
  into the lighter particle $\chi_1$, mediated by the dark photon $A'$.}
\label{fig:Scatter} 
\end{figure}

To make the DM particles $\chi_{1,2}$ interact with the SM particles,
we introduce a kinetic mixing between the dark photon $A'$ and the SM photon. In the basis where the kinetic mixing is removed by the shift of the SM photon, the interactions of $\chi_{1,2}$ and the SM fermions involving one dark photon are given by
\begin{align}
    {\cal L} = g_D A'^{\mu} \left( \chi_1 \partial_\mu \chi_2 -\chi_2 \partial_\mu \chi_1 \right) + \epsilon e A'_{\mu} J_{\rm EM}^\mu ,
\end{align}
where $g_D$ is the gauge coupling constant of the $U(1)_D$, $\epsilon$ is the kinetic mixing parameter,
and $J^{\mu}_{\rm EM}$ is the electromagnetic current.  
Through these interactions, the heavier $\chi_2$ inelastically scatters off an electron
and is converted to the lighter $\chi_1$, as described in Fig.~\ref{fig:Scatter}, 
The small mass difference $\delta$ is converted into electron recoil energy,
which may explain the excess of events observed by the XENON1T experiment.

\section{Inelastic DM-electron scattering}\label{sec:excess}

Let us now evaluate the rate of the inelastic DM electron scattering, taking account of the xenon atomic structure.
For our purpose, we generalize the discussion of
refs.~\cite{Essig:2011nj,Roberts:2019chv,Roberts:2016xfw} into the case of inelastic down-scattering off an electron. 
The differential cross section for a DM velocity $v$ is given by
\begin{align}
\frac{d\sigma v}{d E_R}=\frac{\sigma_e}{2m_e v}\int_{q_-}^{q_+} a_0^2q \, dq K(E_R,q)\ .
\label{differential cross section}
\end{align}
Here, $m_e$ is the electron mass, $a_0=1/(\alpha m_e)$ is the Bohr radius where $\alpha \equiv e^2/4\pi \simeq 1/137$ is the fine structure constant, $E_R$ is the recoil energy, $q$ is the transferred momentum,
$K(E_R,q)$ is the atomic excitation factor, and $\sigma_e$ is the free electron cross section given by
\begin{align}
\sigma_e=\frac{16\pi\epsilon^2\alpha\alpha_D m_e^2}{m_{A'}^4} \; ,
\end{align}
in our model.
Here, $\alpha_D \equiv g_D^2/(4\pi)$ and $m_{A'}$ denotes the dark photon mass.
The limits of the integration $q_\pm$ in Eq.\,\eqref{differential cross section} are determined in the following way.
The energy conservation in the scattering process leads to the relation,
\begin{equation}
\label{eq:energyconservation}
    \frac{q^2}{2m_2}-v q \cos\theta = \delta-E_R  ,
\end{equation}
where $\theta$ is the angle between the momentum of $\chi_2$ and the transferred momentum.
The possible range of $-1 < \cos \theta < 1$ determines the limits of the integration $q_\pm$.
Unlike elastic scattering, we now have two cases of $E_R\geq \delta$ and $E_R \leq \delta$.
For $E_R\geq \delta$, Eq.\,\eqref{eq:energyconservation} leads to
\begin{equation}
\begin{split}
&q_\pm =m_{2} v \pm \sqrt{m_{2}^2 v^2-2m_{2}(E_R-\delta)} \; ,
\label{limits}
\end{split}
\end{equation}
while for $E_R \leq \delta$ we obtain
\begin{equation}
\begin{split}
&q_\pm= \pm m_{2} v+\sqrt{m_{2}^2 v^2-2m_{2}(E_R-\delta)} \; .
\label{limitsnew}
\end{split}
\end{equation}
 Eq.\,\eqref{limits} is reduced to that of elastic scattering in the limit of $\delta \rightarrow 0$, while 
Eq.\,\eqref{limitsnew} is possible only for the case of inelastic down-scattering.

\begin{figure}[!t]
\begin{center}
  \includegraphics[width=7cm]{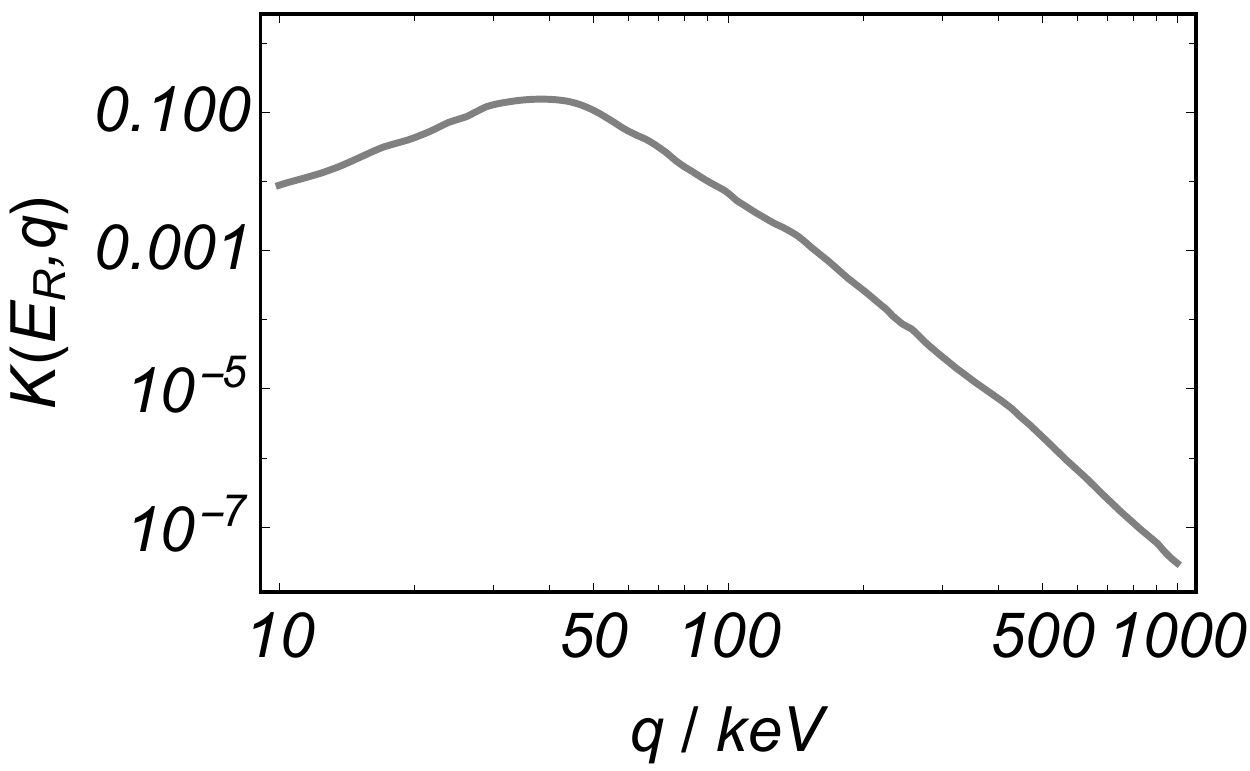}
 \end{center} 
 \vspace{-0.3cm}
  \caption{
The atomic excitation factor $K(E_R,q)$ as a function of the transferred momentum $q$ for $E_R=2\,{\rm keV}$.}
\label{fig:Kfactor} 
\end{figure}

The atomic excitation factor $K(E_R,q)$ is taken from refs.~\cite{Roberts:2016xfw,Roberts:2019chv}.
Its dependence on the momentum transfer $q$ is shown in Fig.~\ref{fig:Kfactor}. Here the contribution from the bound states with the principal quantum number $n=3$ dominates, since their binding energy is around a few keV. The factor is the largest for $q^{-1}$ as large as the size of the wave functions of those states. For elastic scattering, where $\delta = 0$, $q_-$ is at the smallest $O(E_R/v)=O({\rm MeV})$ and the momentum maximizing the atomic factor is irrelevant.
On the other hand, for inelastic down-scattering, when $E_R\sim\delta$, $q_-$ approaches zero and the momentum transfer maximizing the factor is available.

\begin{figure}[!t]
\begin{center}
  \includegraphics[width=7cm]{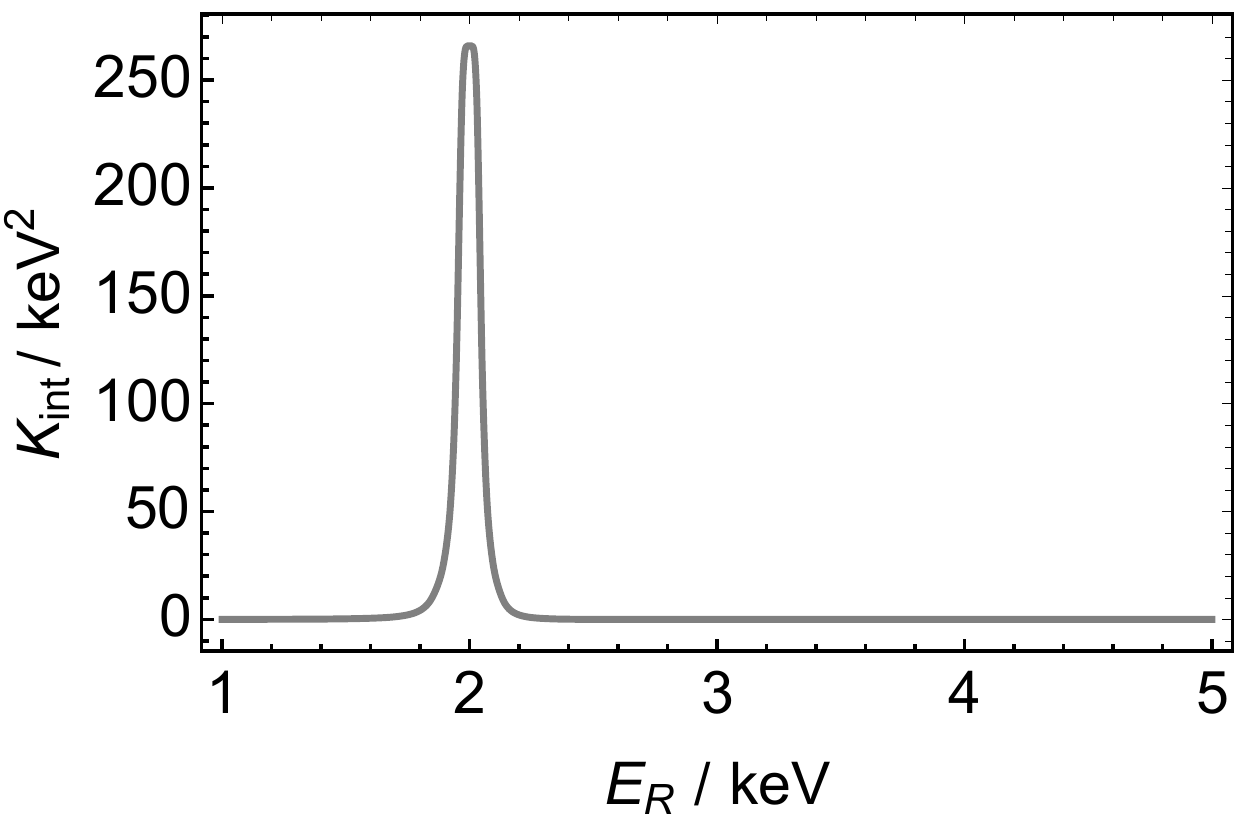}
 \end{center}
 \vspace{-0.3cm}
  \caption{
The atomic excitation factor after the $q$ integration, $K_{\rm int} (E_R)$, defined in Eq.\,\eqref{Kint},
as a function of the transferred recoil energy $E_R$.
Here, we take $m_1=1\,{\rm GeV}$ and $\delta=2\,{\rm keV}$.}
\label{fig:Kint} 
\end{figure}

In Fig.~\ref{fig:Kint}, we plot 
\begin{equation}
 K_{\rm int}(E_R)= \int_{q_-}^{q_+}q\,dq K(E_R,q)\ ,
 \label{Kint}
\end{equation}
as a function of $E_R$ for a representative parameter set of 
$m_1=1\,{\rm GeV}$ and $\delta=2\,{\rm keV}$.
The figure shows a peak around $E_R\simeq\delta$, since $q_-$ approaches zero and the maximal value of $K(E_R,q)$ is available.
We emphasize that this enhancement is a characteristic feature of the inelastic scattering.
The dependence of $K_{\rm int}$ on the DM mass is negligible in the mass range we focus on.

The differential event rate for the inelastic DM scattering with electrons in xenon is given by
\begin{align}
\frac{dR}{dE_R}=n_{\rm T}n_{\chi_2}\frac{d\sigma v}{dE_R}\ ,
\end{align}
where $n_{\rm T} \approx 4\times 10^{27}/{\rm ton}$ is the number density of xenon atoms
and $n_{\chi_2}$ is the number density of the DM component $\chi_2$. We find that $d\sigma v/dE_R$ is almost independent of $v$ around the typical value $10^{-3}$.
The final event rate is given by
\begin{align}
\label{eq:rate}
    R\simeq 4\times 10^9 \epsilon^2g_D^2\left(\frac{1\,{\rm GeV}}{m}\right)\left(\frac{1\,{\rm GeV}}{m_{A'}}\right)^4 \frac{\rho_{\chi_2}}{\rho_{\rm DM}/2}/{\rm ton}/{\rm year} .
\end{align}
Here, $\rho_{\chi_2}$ and $\rho_{\rm DM}$ are the energy density of $\chi_2$ and the total energy density of DM
respectively.
This event rate is about $10^{6}$ times larger than that of elastic scattering per 1 keV bin.

\begin{figure*}
  \begin{minipage}[t]{0.45\hsize}
\centering
\includegraphics[width=7cm]{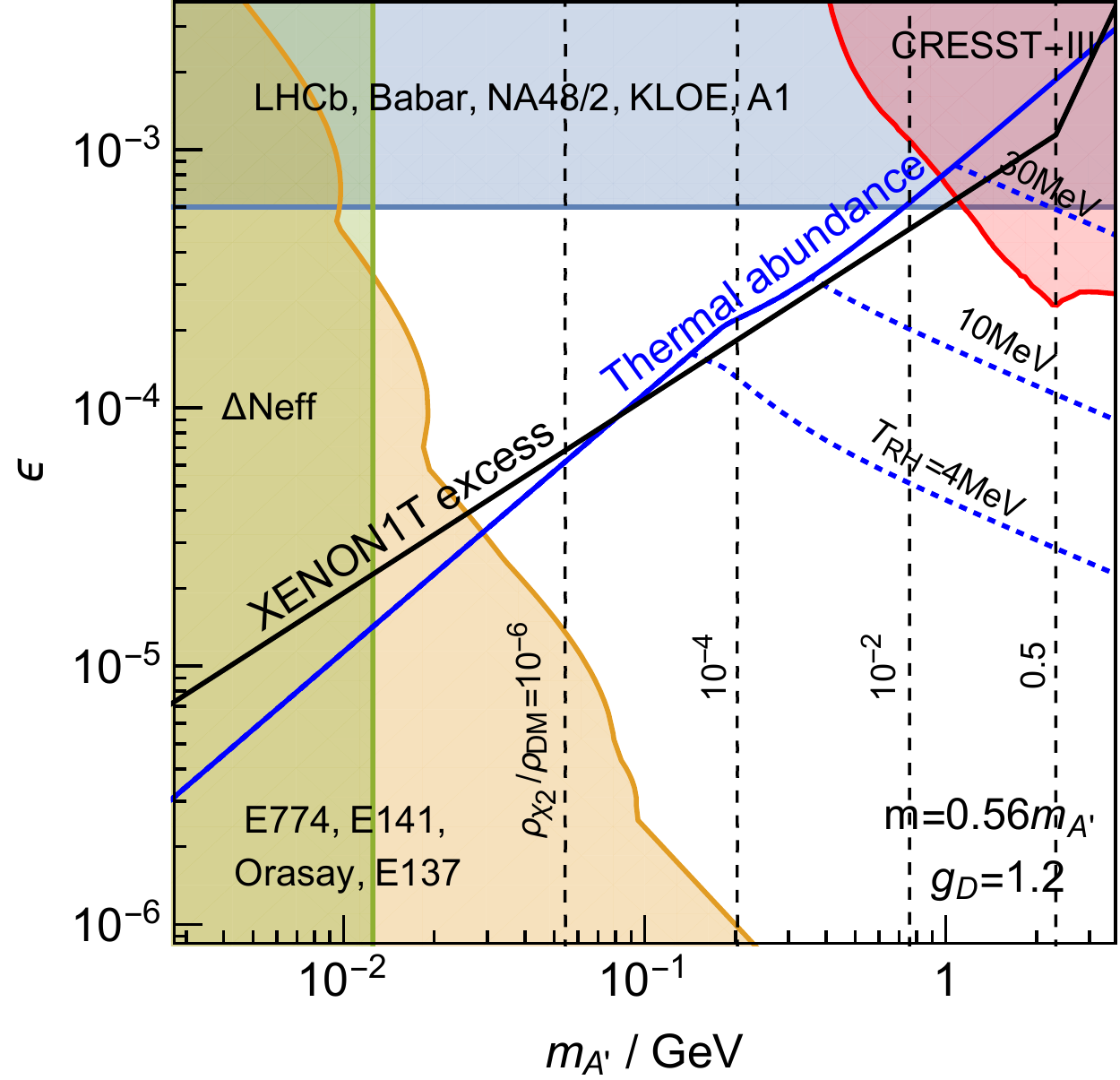}
      \end{minipage}
      \begin{minipage}[t]{0.45\hsize}
   \centering
\includegraphics[width=7cm]{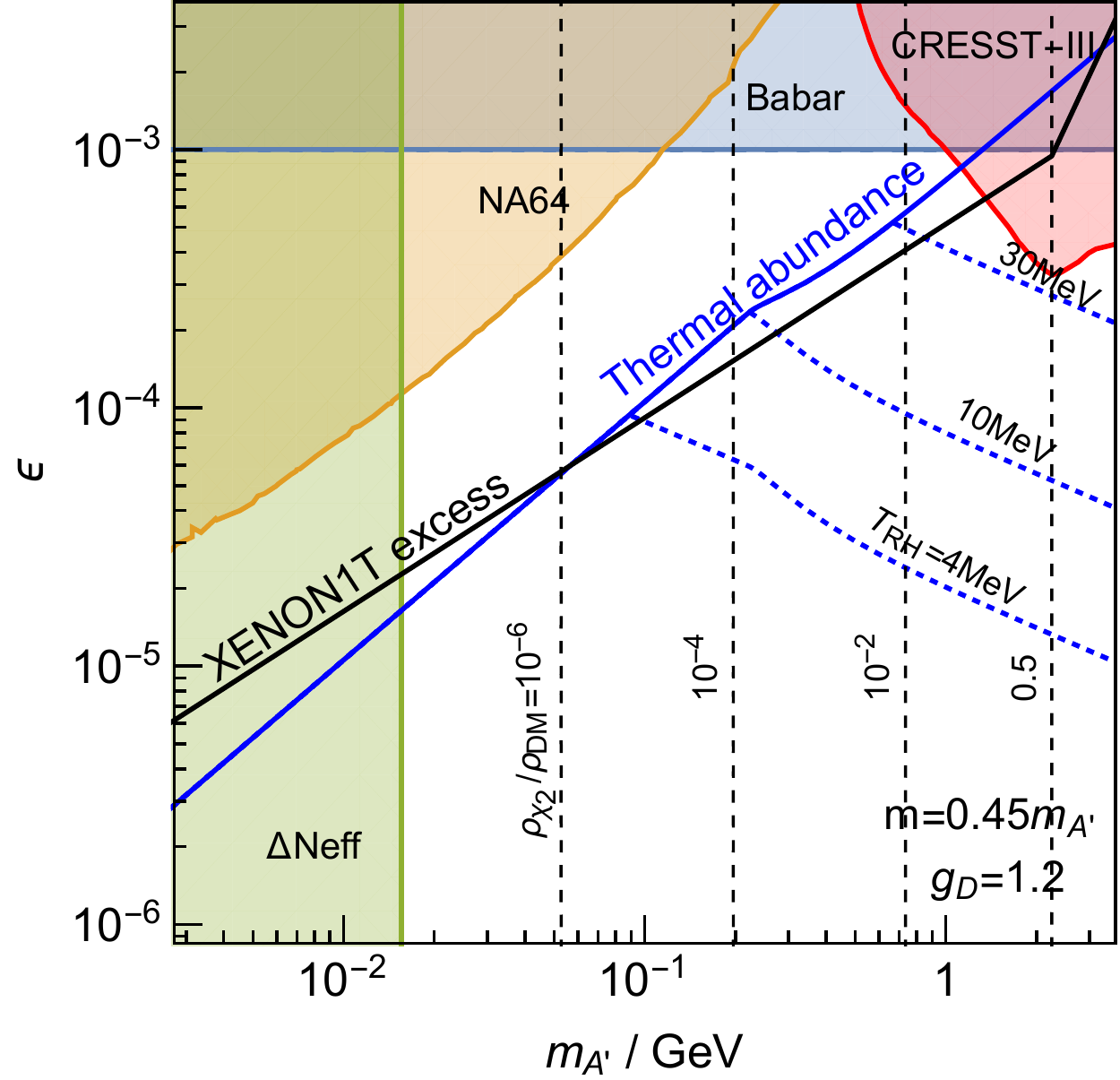}
  \end{minipage}
  \caption{
The required value of $\epsilon$ to explain the observed excess of events at XENON1T
in terms of the dark photon mass $m_{A'}$ (black solid lines).
The left and right panels correspond to the cases of $m > m_{A'}/2$ and $m < m_{A'}/2$ respectively.
We assume $g_D =1.2$ in both cases.
The blue lines denote the required value of $\epsilon$ to obtain the observed DM abundance by the thermal freeze-out process, discussed in Sec.~\ref{sec:FO}. The solid lines correspond to the case without any entropy production.
The dashed lines assume freeze-out during a matter dominated era and the subsequent reheating at $T_{\rm RH}$, which suppresses the DM abundance by a factor of $\left(T_{\rm RH}/T_{\rm FO}\right)^3$.
The black dashed lines denote the mass density of $\chi_2$ normalized by the total DM density.
The shaded regions show the constraints from dark radiation and various searches for the dark photon $A'$
which are discussed in Sec.~\ref{sec:constraints}.
}
\label{fig:mchi06mAD} 
\end{figure*}

\section{The Relic abundance}
\label{sec:FO}

In this section, we discuss the abundance of the DM particles $\chi_{1,2}$ produced by the thermal freeze-out process.
Since the mass difference $\delta$ is much smaller than the freeze-out temperature $T_{\rm FO} \sim m/10$,
we may start with using the complex scalar field $\phi$ to compute the abundance of $\chi_{1,2}$.

To evade the direct detection bound from the nuclear recoil experiments, $m$ must be below the GeV scale. For such a small mass, the CMB constraint excludes the thermal freeze-out production of DM determined by s-wave annihilation
\cite{Padmanabhan:2005es,Aghanim:2018eyx}.
We thus consider the case where the relic abundance is determined by the p-wave annihilation of $\phi$ to a pair of SM fermions through the s-channel exchange of $A'$.
The annihilation cross section of $\phi$ into a fermion $f$ is given by
\begin{align}
    \sigma v \simeq \frac{\epsilon^2 q_f^2 e^2 g_D^2}{6\pi} \frac{m^2}{m_{A'}^4} v^2 \times \frac{1}{\left(1 - 4m^2/m_{A'}^2\right)^2},
\end{align}
where $v$ is the relative velocity of the initial states and $q_f$ is the electric charge of the fermion $f$.
The correct relic abundance of DM is obtained for
\begin{equation}
\begin{split}
    \epsilon g_D \simeq 3\times 10^{-4} \times \left(\frac{m_{A'}}{0.1 \, {\rm GeV}} \right)^2 \left( \frac{0.1 \, {\rm GeV}}{m} \right) \\
    \times\left|1 - \frac{4m^2}{m_{A'}^2} \right| \left(\sum_{f} {q_f^2}\right)^{-1/2},
\end{split}
\end{equation}
where the summation over $f$ is taken for the SM fermions lighter than $\phi$. For $m_{A'}$ near $2m$, the required value of $\epsilon g_D$ is suppressed because of the enhancement of the annihilation by the dark photon pole. The blue lines in Fig.~\ref{fig:mchi06mAD} show the required value of the kinetic mixing $\epsilon$ to explain the observed DM abundance
by the thermal freeze-out of $\chi_{1,2}$.

Although the total number density of $\chi$ is conserved after the freeze-out of the annihilation into SM fermions, $\chi_{1,2}$ continue to be converted to each other by scatterings with the SM particles and more importantly the scattering $\chi_1\chi_1 \leftrightarrow \chi_2 \chi_2$
\cite{Finkbeiner:2009mi,Batell:2009vb} (see also ref.~\cite{Baryakhtar:2020rwy})%
\footnote{The scattering among $\chi$ was missed in the first version of this paper. We thank Masha Baryakhtar, Asher Berlin, Hongwan Liu and Neal Weiner for pointing out this process to us.}
whose cross section is
\begin{align}
    \sigma v = \frac{g_D^4}{\pi} \frac{m}{m_{A'}^4}\sqrt{2m \delta} \; .
\end{align}
Once the temperature drops below the electron mass, electrons almost disappear from the thermal bath, and scatterings with the SM particles freeze-out. Scatterings with the asymmetric component of electrons as well as with photons are negligible. The temperature of $\chi$ begins to decrease in proportion to the square of the scale factor of the universe. When the temperature of $\chi$, $T_\chi \simeq T_\gamma^2 / m_e$, drops below the mass difference $\delta$, the number density of $\chi_2$ begins to be exponentially suppressed, and the process $\chi_1\chi_1 \leftrightarrow \chi_2 \chi_2$ eventually freezes-out. The mass density of $\chi_2$ normalized by the DM density is
\begin{align}
\label{eq:fraction}
    \frac{\rho_{\chi_2}}{\rho_{\rm DM}} \simeq \frac{0.04}{g_D^4}  \left(\frac{m_{A'}}{{\rm GeV}}\right)^4 \left(\frac{{\rm GeV}}{m}\right)^{1/2} \frac{2~{\rm keV}}{\delta} \left(\frac{\delta/T_{\chi,\rm FO}}{5}\right)^{1/2},
\end{align}
where $T_{\chi,{\rm FO}}$ is the temperature of $\chi$ when the scattering freeses-out. Here it is assumed that the scattering is effective when $T_\chi \sim \delta$.  Otherwise, the relic abundances of $\chi_{1,2}$ are identical, $\rho_{\chi_2} = \rho_{\rm DM}/2$.



\section{(In)direct constraints}\label{sec:constraints}

\begin{figure}[!t]
\begin{center}
  \includegraphics[width=6cm]{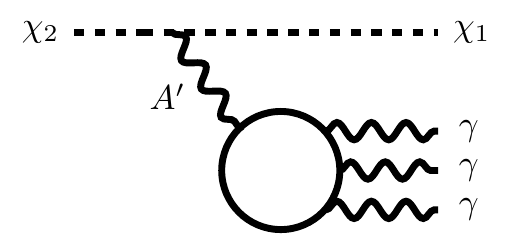}
 \end{center} 
  \vspace{-0.3cm}
  \caption{The decay of the heavier DM particle $\chi_2$ into the lighter $\chi_1$ and three photons,
  mediated by the dark photon $A'$ and the SM fermion loop.}
\label{fig:decay} 
\end{figure}


We here discuss (in)direct constraints on the model parameter space.
Since the p-wave annihilation of the DM particles $\chi_{1,2}$ at a low temperature is suppressed,
constraints from the CMB and the indirect detection experiments are not relevant in our model. For $m_{A'}/2< m < m_{A'}$, $\chi_{1,2}$ annihilates into $A'$ and SM particles through an off-shell $A'$, but we find that this does not constrain the parameter region explaining the XENON1T excess.

The DM-nucleon scattering cross section is
\begin{align}
    \sigma_n =& \,\, 4\alpha \epsilon^2 g_D^2 \left(\frac{Z}{A}\right)^2 \frac{m_N^2}{m_{A'}^4} \nonumber \\
    \simeq & \,\, 6\times 10^{-38}{\rm cm}^2 \left(\frac{Z/A}{0.5}\right)^2 \frac{m}{\rm GeV} \frac{R}{100/{\rm ton}/{\rm year}},
\end{align}
where $Z$ is the atomic number, $A$ is the atomic weight, and $m_N$ is the nucleon mass. In the second line we expressed the parameters in terms of the signal rate at XENON1T. The bound from CRESST-III~\cite{Abdelhameed:2019hmk} is satisfied for $m\lesssim 0.7$ GeV. Here we simply scaled the constraint on the elastic scattering by the fraction $\rho_{\chi_2}/\rho_{\rm DM}$. The recoil energy for inelastic scattering is typically around two times larger than that of elastic scattering, so the actual upper bound on $m$ may be few times stronger.

The heavier state $\chi_2$ can decay into the lighter state $\chi_1$ and three photons via the diagram shown in Fig.~\ref{fig:decay}.
The decay rate is roughly given by
\begin{align}
    \Gamma\sim \frac{1}{(4\pi)^5}\alpha^4 \epsilon^2 g_D^2 \frac{\delta^{13}}{m_{A'}^4m_e^8},
\end{align}
which is much smaller than the upper bound from the X-ray search (see e.g.~\cite{Adhikari:2016bei})
in the parameter space we consider.
Note that the decay of $\chi_2$ into $\chi_1$ and one or two photons is highly suppressed.
Then, any constraint from the decays of $\chi_2$ is negligible.  

When $m$ is small, the annihilation of $\chi$ after neutrinos decouple heats up only electrons and photons, leading to a negative amount of dark radiation $\Delta N_{\rm eff}$. Using the results in~\cite{Boehm:2013jpa} and the latest constraint on $\Delta N_{\rm eff}$ from the Planck observation~\cite{Aghanim:2018eyx}, we find $m>7$ MeV.

The parameter space is further constrained by the dark photon searches.
For $m > m_{A'}/2$, the dark photon $A'$ does not decay into $\chi_{1,2}$ and decays only visibly,
so that the direct search constraint is rather strong.
The left panel of Fig.~\ref{fig:mchi06mAD} shows the constraints from the searches at LHCb~\cite{Aaij:2017rft}, Babar~\cite{Lees:2014xha}, NA48/2~\cite{Batley:2015lha}, KLOE~\cite{Archilli:2011zc,Babusci:2012cr,Babusci:2014sta,Anastasi:2016ktq}, A1~\cite{Merkel:2014avp}, E774~\cite{Bross:1989mp}, E141~\cite{Riordan:1987aw}, Orasay~\cite{Davier:1989wz}, and E137~\cite{Bjorken:1988as,Batell:2014mga}, as summarized in ref.~\cite{Beacham:2019nyx}.
For $m < m_{A'}/2$, since $A'$ can decay into $\chi_{1,2}$, the constraints tend to be relaxed,
as shown in the right panel of the figure.
Here we show the constraints from Babar~\cite{Lees:2017lec} and NA64~\cite{Banerjee:2016tad}.

\section{The XENON1T excess}\label{xenon1t}

In Fig.~\ref{fig:mchi06mAD}, using Eqs.~(\ref{eq:rate}) and (\ref{eq:fraction}), we show the value of the kinetic mixing $\epsilon$ which explains the excess of events observed at XENON1T with a rate $\approx 100$/ton/year by the black solid lines. We can see that the observed data is fitted by a wide range of the kinetic mixing
for both the cases of $m > m_{A'}/2$ and $m < m_{A'}/2$.

There exists a parameter set which explains the total abundance of $\chi_{1,2}$ by the thermal freeze-out as well as the XENON1T. For larger $g_D$, the black lines go up since the relative abundance of $\chi_2$ decreases, while the blue lines go down. The blue and black lines cross at higher $m_{A'}$,
For smaller $g_D$, they cross at lower $m_{A'}$. As a result, the crossing point exists in a viable parameter region when $1< g_D<1.4$ for $m= 0.56\,m_{A'}$, and $1< g_D<1.5$ for $m= 0.45\,m_{A'}$.
For $m$ closer to $m_{A'}/2$, since the annihilation cross section of $\chi_{1,2}$ into SM particles becomes larger, the blue lines go down. The blue and black lines cross at higher $m_{A'}$. The viable range of $g_D$ shifts downward.
The simultaneous explanation of the abundance of $\chi_{1,2}$ by the thermal freeze-out and the XENON1T excess requires large $g_D$ and/or $m$ close to $m_{A'}/2$.
Otherwise, after fixing the parameters to explain the XENON1T excess, the abundance of $\chi_{1,2}$ determined by the thermal freeze-out is too large.

The overproduction may be avoided by some entropy production. If the freeze-out occurs during a matter dominated era, the abundance of $\chi_{1,2}$ is suppressed by a factor of $\left(T_{\rm FO}/T_{\rm RH}\right)^3$
where $T_{\rm RH}$ is the reheating temperature.
Fig.~\ref{fig:mchi06mAD} shows the required value of $\epsilon$ for several $T_{\rm RH}$.
The BBN constraint requires $T_{\rm RH} > 4 \, \rm MeV$
\cite{Kawasaki:1999na,Kawasaki:2000en}. A similar bound is also obtained from the CMB~\cite{deSalas:2015glj},
since for a lower $T_{\rm RH}$ the reheating after neutrinos decouple is non-negligible
and the neutrinos become relatively cooler than photons. The figure
indicates that there exists a parameter region which simultaneously explains the XENON1T excess
and realizes the correct DM abundance via the thermal freeze-out process and entropy production. 

Another possibility to change the relic abundance is to introduce an additional particle
to which $\chi_{1,2}$ annihilate.
If such a particle is massless, it behaves as dark radiation. The massless new particle decouples from the thermal bath around when $\chi_{1,2}$ decouple. If this occurs after the QCD phase transition, the massless particle contributes to too much dark radiation~\cite{Green:2019glg}. To avoid this case, $m \gtrsim$ GeV is required, which is excluded by nuclear recoil experiments discussed in section~\ref{sec:constraints}. If the new particle is massive,
it may decay into the SM particles.
In this case, the annihilation should not contain an s-wave. One viable example is the annihilation of $\chi_{1,2}$ into a new scalar particle which is charged under the $U(1)_D$ and mixes with the SM Higgs boson
after the electroweak and $U(1)_D$ symmetry breaking.

\section{Conclusion}\label{sec:discussion}

We have proposed a new interpretation of the electron recoil excess observed in the XENON1T experiment
where one of the DM components $\chi_2$ inelastically scatters off an electron and is converted into the lighter DM component $\chi_1$. 
The mass difference between $\chi_1$ and $\chi_2$ is converted into electron recoil energy.
The DM-electron interaction is mediated by a massive dark photon $A'$.
The lifetime of the heavier $\chi_2$ is sufficiently long and the X-ray search does not give a constraint on
the relevant parameter space.
We evaluated the rate of the inelastic DM scattering, taking account of the xenon atomic structure,
and found a viable parameter space which explains the observed excess
and is consistent with various dark photon searches.
Some viable parameter region is consistent with the DM relic abundance by the thermal freeze-out process.
Even for the other region, some entropy production or a new annihilation mode may be able to address the issue.

In the evaluation of the inelastic DM electron scattering rate, taking account of the atomic structure,
we found that the scattering rate is significantly enhanced for a recoil energy at the mass difference, since the momentum transfer is allowed to be small and can match the size of the wave function of the electrons in the atom. This feature is not seen for elastic scattering and
it is worth studying it in more general setups for applications to other direct detection experiments.

Various experiments searching for dark photons are ongoing and also proposed
(see ref.~\cite{Beacham:2019nyx} and references therein).
It is important to explore future prospects
to test our model parameter space which can explain the XENON1T electron recoil excess. 

\section*{Acknowledgements}
We would like to thank Benjamin Roberts and Yevgeny Stadnik for discussions.
Y.N.~and M.S.~are grateful to Kavli IPMU for their hospitality during the COVID-19 pandemic.
This work was supported in part by the DoE grants DE-SC0009988 (K.H.) as well as the Raymond and Beverly Sackler Foundation Fund (K.H.).

\bibliography{bib}
\bibliographystyle{utphys}

\end{document}